# Synthesis, Structural, Electrical and Microchemistry Properties of Novel (LSCFM) $La_xSr_{1-x}Co_{1-y-z}Fe_yMn_zO_{3-\delta}$ (x=0.4, 0.6, 0.8, y=0.4, 0.8 and z=0.1, 0.2) Perovskites for SOFC Applications

E. A. Tsiligkeridis[1], S. S. Makridis[1,] *, E. Pavlidou[2], E. S. Kikkinides[1] and A. K. Stubos[3]

[1] Department of Mechanical Engineering, University of Western Macedonia, Kozani, GR 50100, Greece

[2] Department of Physics, Aristotle University of Thessaloniki, GR 54124, Greece

[3] Institute of Nuclear Technology and Radiation Protection, NCSR "Demokritos", Athens, GR 15310, Greece

Keywords: composite materials; solid state reactions; crystal structure; Perovskite oxide; Solid oxide fuel cell (SOFC); microstructure

## Abstract

Perovskites are mixed oxides with the general formula $ABO_3$. Many different compositions have been found to be possible and their number is growing rapidly due to their characteristics that make them ideal for various applications, such as fabrication of cathodes for solid oxide fuel cells (SOFCs). Oxide powders of $La_xSr_{1-x}Co_{1-y-z}Fe_yMn_zO_{3-\delta}$ (x=0.4, 0.6, 0.8, y=0.4, 0.8 and z=0.1, 0.2), named LSCFM, were prepared by mixing commercial powders. The powders have been investigated as a function of the stoichiometry and the crystal structure parameters have been determined using Rietveld analysis. The results of the structural analysis for each sintered sample, at low temperature sintering (1100 °C), have been obtained by considering two cubic perovskite phases, for x=0.4 or one cubic and one orthorhombic phase for x=0.6, 0.8, respectively. Analysis of microstructure has shown a maximum grain size of ~0.63 μm. The annealed sample at 1100 °C, an almost single phase specimen, exhibited an electrical conductivity of $1.29 \times 10^3$ $(\Omega \cdot cm)^{-1}$. The sintered sample at 1250 °C has higher conductivity but sharper curve close to the maximum. We suggest that low temperature sintering reveal two phase system with broader curve maximum and high conductivity for future applications in SOFCs.

## 1. Introduction

In recent years, compositions in the $(La,Sr)(Co,Fe)O_{3-\delta}$ system ($ABO_3$ perovskite structure) have been investigated because of their mixed conducting behavior. At elevated temperatures, these solid-solution compositions exhibit substantial ionic and electronic conductivity, which makes them attractive candidate materials for several important applications, including solid oxide fuel cell cathodes, oxygen separation membranes, and membrane reactors for syngas production and the partial oxidation of hydrocarbons. It is important for these solids that they maintain their electro-neutrality and that the A and B cations be stable.

The ideal oxide perovskite structure $ABO_3$ consists of a cubic array of corner sharing $BO_6$ octahedra with the A cation at the body centre position. However non-ideal A and B ionic radii create distortions to the cubic lattice (often orthorhombic or rhombohedral) [1-3].

Perovskite-type strontium doped lanthanum manganites, $La_{1-x}Sr_xMnO_{3-\delta}$ have attracted considerable interest for a long time because of their structural [4, 5], electronic [4] and magnetic properties [5, 6]. However, the high operating temperatures (800 to 1000) °C of perovskite-type strontium doped lanthanum manganites, cause economic and technical difficulties associated with material property requirements. Therefore, research is directed towards reducing these operating temperatures.

At intermediate temperatures (600 to 800) °C it has been found that the kinetics of the cathodic reaction reduces cell performance [7]. Thus, optimization of SOFC electrode materials and their performance is essential. Good candidates for the cathode are $LaCoO_3$-based perovskite oxides, and particularly those substituted with Sr and Fe on A and B sites, respectively, else known as LSCFs.

These materials are compatible with ceria-gadolinia electrolytes and have suitable stability ranges, while possesing good conductivity at intermediate temperatures [8]. Furthermore, the surface oxygen exchange and bulk diffusion characteristics are superior to those of conventional strontium-doped lanthanum manganite (LSM) cathodes [9].

Badwal et al. [10] have claimed that due to their mixed conductivity these materials exhibited two important functions; charge transport through both ions and electronic carriers and efficient charge transfer at the oxygen/electrode interface.

The defect perovskite $LaMnO_{3+\delta}$ has shown a structure transformation from orthorhombic to cubic (via rhombohedral) with increase of $Mn^{4+}$ content. Cell dimensions and characteristics depend not only on the $Mn^{4+}$ content but also on the thermal treatment [4]. As the Mn concentration increases, direct transport among Mn sites becomes possible and the conductivity increases [11].

Since LSCFs tend to prevail as ideal composites for cathodes in SOFCs, the effect of manganese (Mn) should be examined as a substitution for B site. The aim is to affect favorably the structural properties which results in higher conductivity values.

## 2. Experimental

The specimens in this study have been prepared by the conventional powder mixing method. The oxide powders of $La_xSr_{1-x}Co_{1-y-z}Fe_yMn_zO_{3-\delta}$ (x=0.4, 0.6, 0.8, y=0.4, 0.8 and z=0.1, 0.2) have been synthesized by blending commercial powders of $La_2O_3$ (Alfa Aesar, 99.99 % pure), SrO (Alfa Aesar), CoO (Alfa Aesar, 95 % pure), $Fe_3O_4$ (Alfa Aesar, 97 % pure), and MnO (Alfa Aesar, 99 % pure) in stoichiometric ratio. The starting powders have been mixed-milled in an achate mortar and pestle, then annealed at 1100 °C and 1350 °C for 8 days followed by slow cooling at room temperature. The samples have been pressed into pellets with a hydraulic press (Mega KSC-15A) at 44 MPa.

The crystal structure of the ceramics has been investigated as a function of the stoichiometry. X-ray diffraction patterns have been obtained at room temperature by using a Seifert XRD 3003 TT diffractometer with Cu-$K_\alpha$ ($\lambda_{K\alpha1}$=1.54186 Å) radiation, scanning rate 0.02 °·s$^{-1}$ and a Siemens D-5000 X-ray diffractometer. Structural characteristics have been determined by X-ray Rietveld analysis on the RIETICA structure refinement program.

The morphology and the microchemistry of the pellets have been characterized by scanning electron microscopy (SEM) and by energy dispersive analysis of X-Rays (EDAX), using a 20 kV JEOL 840A microscope equipped with an OXFORD ISIS 300 EDS analyzer and the necessary software. Microanalysis and chemical mapping have been performed along the surface of the pellets. An average grain size has been obtained by using the Scion Image software. The sub-micron powder has been uniaxially pressed into rectangular bar (30 mm × 4 mm × 4 mm), followed by sintering at 1100 $^oC$ for 4 h in air. The rectangular specimen has been painted with platinum paste for measuring electrical conductivity. The latter has been determined at 50 $^oC$ to 950 $^oC$ by a dc four-terminal method in air.

## 3. Results and discussion

The initially obtained X-ray diffraction patterns by the Seifert XRD were inadequate for an accurate investigation of the structure. Thus, better quality XRD measurements were performed on all samples using a Siemens 5000D diffractometer. The XRD patterns acquired with the two diffractometers are compared in Fig. 1.

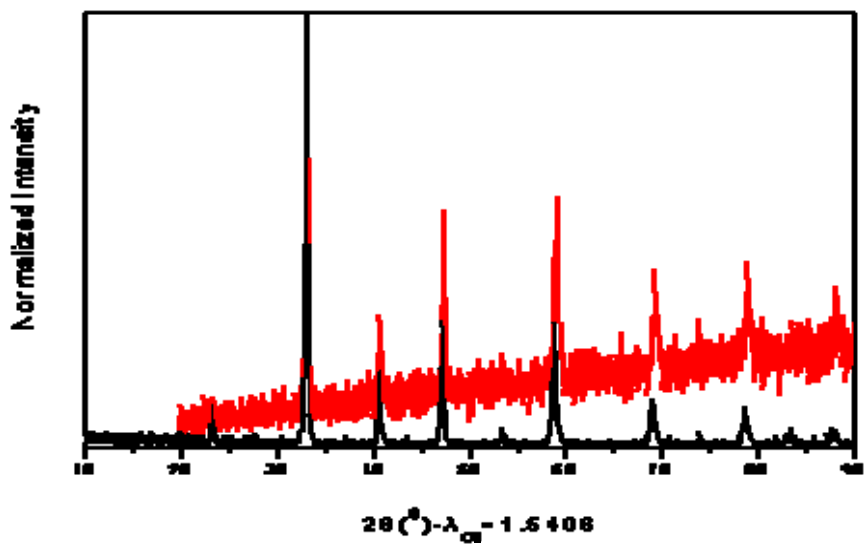

**Figure 1.** X-ray diffraction pattern produced by Siefert 3003TT XRD (shown in red) and by Siemens 5000D (shown in black) equipped with Si-monochromator.

The comparison of the theoretical patterns of R-3C and PM3M structures, produced by RIETICA software, does not provide a clear picture of the LSCFMs crystal structure (Fig.2).

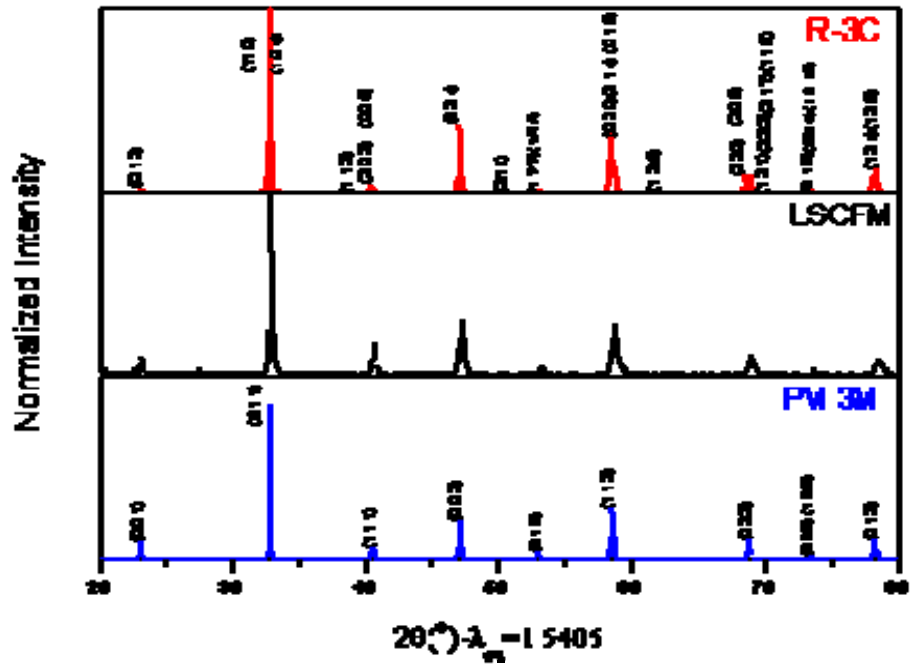

**Figure 2.** Comparison of the theoretical R-3C and PM3M phase structures.

The results of the Rietveld analysis for each sample revealed two cubic perovskite phases for x=0.4 (Fig.3) and one cubic and one orthorhombic phase for x=0.6, and 0.8, respectively (Fig. 4). As found in the analysis, the two cubic phases in Fig. 3 have almost the same lattice parameters while their mass percentages are 55 % and 45 % respectively. This quantitative analysis is directly connected to the microstructure investigation where the back scattered electron analysis revealed two different phases through the luminance gradient.

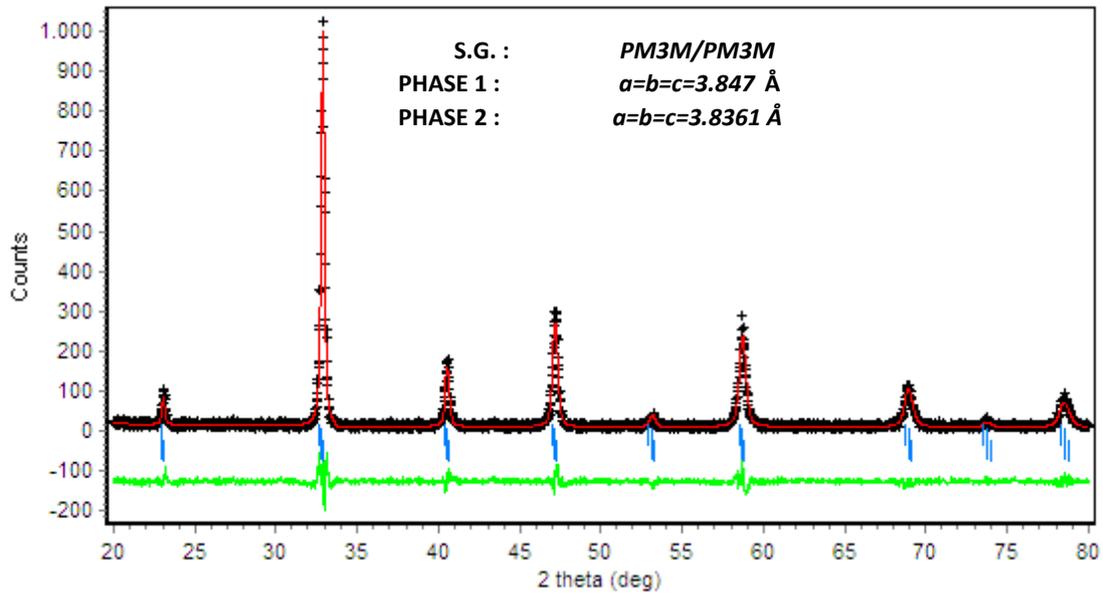

**Figure 3.** X-ray diffraction pattern and Rietveld analysis of $La_{0.4}Sr_{0.6}Co_{0.4}Fe_{0.4}Mn_{0.2}O_{3-\delta}$ perovskite, by using two cubic structures.

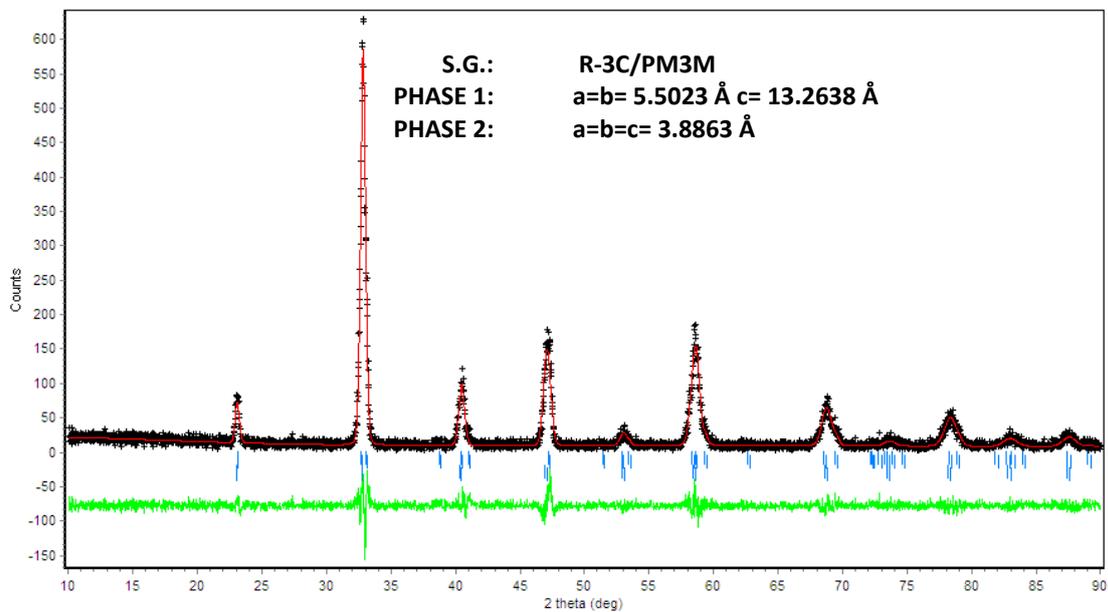

**Figure 4.** X-ray diffraction pattern and Rietveld analysis of $La_{0.6}Sr_{0.4}Co_{0.4}Fe_{0.4}Mn_{0.2}O_{3-\delta}$ perovskite, by using one rhombohedral and one cubic structure.

The results of the Rietveld analysis of $La_xSr_{1-x}Co_{1-y-z}Fe_yMn_zO_{3-\delta}$ (x=0.4, 0.6, 0.8, y=0.4, 0.8 and z=0.1, 0.2) perovskites are shown in Table 1. It can be seen that complete homogenization is not accomplished for x=0.4. As the La content increases, the LSCFMs tend to crystallize in a single structure, as revealed by the Rietveld analysis. The complete homogenization of the microstructure and single phase perovskite structure, as was expected, developed after thermal treatment at 1250 °C. After sintering, there was no evidence of diffraction peaks corresponding to the parent oxides. From the reliability factors and the estimated probable errors it reveals that the cubic and the orthorhombic phases could be present together.

Table 1.
Summary of Rietveld analysis of $La_xSr_{1-x}Co_{1-y-z}Fe_yMn_zO_{3-\delta}$ (x=0.4, 0.6, 0.8, y=0.4, 0.8 and z=0.1, 0.2) perovskites.

| Nominal Composition | Rietveld Refinement | | | | | weight % of phases | | Crystal Structure Parameters (a,b,c in Å) | |
|---|---|---|---|---|---|---|---|---|---|
| | Space group | $R_{bragg}$ | $R_p$ | $R_{wp}$ | $\chi^2$ | Phase 1 | Phase 2 | Phase 1 | Phase 2 |
| $La_{0.4}Sr_{0.6}Co_{0.1}Fe_{0.8}Mn_{0.1}O_{3-\delta}$ | Double Cubic | 2.89 | 13.00 | 16.77 | 1.157 | 75.63 | 24.37 | a=b=c= 3.8669 | a=b=c= 3.8870 |
| $La_{0.4}Sr_{0.6}Co_{0.4}Fe_{0.4}Mn_{0.2}O_{3-\delta}$ | Double Cubic | 4.28 | 16.51 | 23.09 | 1.358 | 55.13 | 44.87 | a=b=c= 3.8470 | a=b=c= 3.8361 |
| $La_{0.4}Sr_{0.6}Co_{0.5}Fe_{0.4}Mn_{0.1}O_{3-\delta}$ | Double Cubic | 3.53 | 11.61 | 14.75 | 1.093 | 72.22 | 27.78 | a=b=c= 3.8542 | a=b=c= 3.8465 |
| $La_{0.6}Sr_{0.4}Co_{0.1}Fe_{0.8}Mn_{0.1}O_{3-\delta}$ | R-3C/ PM3M | 3.60 | 15.47 | 11.71 | 0.036 | 55.54 | 44.46 | a=b= 5.4906 c= 13.3512 | a=b=c= 3.9059 |
| $La_{0.6}Sr_{0.4}Co_{0.4}Fe_{0.4}Mn_{0.2}O_{3-\delta}$ | R-3C/ PM3M | 2.77 | 17.53 | 23.78 | 1.217 | 19.51 | 80.49 | a=b= 5.5023 c= 13.2638 | a=b=c= 3.8863 |
| $La_{0.6}Sr_{0.4}Co_{0.5}Fe_{0.4}Mn_{0.1}O_{3-\delta}$ | R-3C/ PM3M | 2.35 | 11.86 | 15.05 | 1.117 | 99.06 | 0.94 | a=b= 5.4504 c= 13.4292 | a=b=c= 3.8730 |
| $La_{0.8}Sr_{0.2}Co_{0.1}Fe_{0.8}Mn_{0.1}O_{3-\delta}$ | R-3C/ PM3M | 5.08 | 14.77 | 18.67 | 1.201 | 2.95 | 97.05 | a=b= 5.2283 c= 15.3420 | a=b=c= 3.9100 |
| $La_{0.8}Sr_{0.2}Co_{0.4}Fe_{0.4}Mn_{0.2}O_{3-\delta}$ | R-3C/ PM3M | 3.26 | 19.62 | 15.45 | 0.031 | 98.47 | 1.53 | a=b= 5.4768 c= 13.2583 | a=b=c= 3.8900 |

Scanning electron microscopy (SEM) images are shown in Fig. 5 while the energy dispersive analysis of X-rays (EDAX) has confirmed the existence of two crystalline phases (perovskites). EDAX chemical analysis of $La_{0.4}Sr_{0.6}Co_{0.1}Fe_{0.8}Mn_{0.1}O_{3-\delta}$ reveals a grey region in the micrograph which is Mn-Co-Fe-La free and a white region having normal atomic composition. In the case of $La_{0.8}Sr_{0.2}Co_{0.4}Fe_{0.4}Mn_{0.2}O_{3-\delta}$ the white region is Mn-Fe-Co-Sr poor while the grey region has less La atomic percentage. The crucial result of the above analysis is that the composite samples have different microchemistry depending on the crystal structure symmetry.

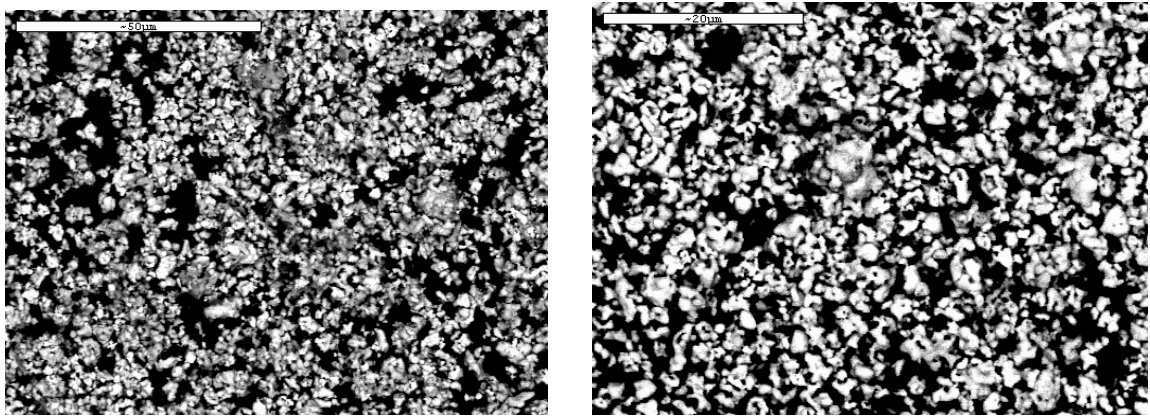

**Figure 5.** Micrographs of scanning electron microscopy of $La_{0.4}Sr_{0.6}Co_{0.1}Fe_{0.8}Mn_{0.1}O_{3-\delta}$ (left) and that of $La_{0.8}Sr_{0.2}Co_{0.4}Fe_{0.4}Mn_{0.2}O_{3-\delta}$ (right) perovskites.

It is well known that when the particle size becomes smaller, the magnetization decreases markedly and subsequently an efficient charge transfer results at the oxygen/electrode interface.

Analysis of scanning electron microscopy pictures using Scion Image software, shown in Fig. 6, revealed that the maximum particle size is 0.63 μm. The grain size is connected to the activated surface and consequently affects significantly the electrical conductivity and the charge transfer kinetics. The calculated grain size from the analysis of the micrographs, captured by scanning electron microscopy, is less than the previous manganese free LSCF perovskites [12]. Table 2 shows the average grain size in μm for all LSCFM materials.

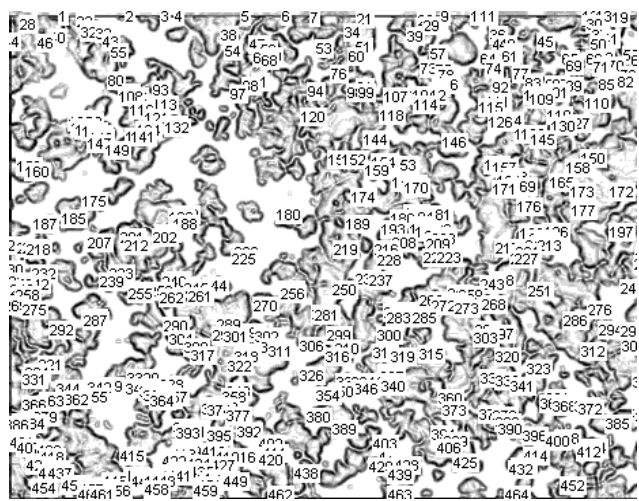

**Figure 6.** Analysis of SEM image for $La_{0.8}Sr_{0.2}Co_{0.5}Fe_{0.4}Mn_{0.1}O_{3-\delta}$ perovskite, with Scion Image Software [13].

Table 2.

Average grain size of LSCFM perovskites after SCION image analysis

| $La_xSr_{1-x}Co_{1-y-z}Fe_yMn_zO_{3-\delta}$ | Average Grain Size (μm) |
|---|---|
| $La_{0.4}Sr_{0.6}Co_{0.1}Fe_{0.8}Mn_{0.1}O_e$ | 0.630939 |
| $La_{0.4}Sr_{0.6}Co_{0.4}Fe_{0.4}Mn_{0.2}O_e$ | 0.131268 |
| $La_{0.4}Sr_{0.6}Co_{0.5}Fe_{0.4}Mn_{0.1}O_e$ | 0.206304 |
| $La_{0.6}Sr_{0.4}Co_{0.1}Fe_{0.8}Mn_{0.1}O_e$ | 0.435156 |
| $La_{0.6}Sr_{0.4}Co_{0.4}Fe_{0.4}Mn_{0.2}O_e$ | 0.293030 |
| $La_{0.8}Sr_{0.2}Co_{0.1}Fe_{0.8}Mn_{0.1}O_e$ | 0.206029 |
| $La_{0.8}Sr_{0.2}Co_{0.4}Fe_{0.4}Mn_{0.2}O_e$ | 0.222551 |
| $La_{0.8}Sr_{0.2}Co_{0.5}Fe_{0.4}Mn_{0.1}O_e$ | 0.176557 |

Fig. 7 shows the electrical conductivity (denoted as $\sigma_e$) of $La_xSr_{1-x}Co_{1-y-z}Fe_yMn_zO_{3-\delta}$ ceramics as a function of temperature. Due to the low oxygen ionic transport number in the $La_{1-x}Sr_xCo_{1-y}Fe_yO_{3-\delta}$ oxides (generally less than 1 %), the electrical conductivity measured by the dc four-terminal method can be regarded as representative of electronic conductivity [14]. The electrical conductivities of the specimens sintered at different temperatures display an identical variation with temperature; the respective values increase first with temperature up to a maximum value near 600 °C and then decrease. The specimen sintered at this temperature exhibited an electrical conductivity of $1.29 \times 10^3$ $(\Omega \cdot cm)^{-1}$ which is a very interesting result for future applications in SOFC. The $La_{0.6}Sr_{0.4}Co_{0.8}Fe_{0.2}O_{3-\delta}$ sample sintered at 1100 °C is characterized by a slightly lower value of electrical conductivity [12]. As we observed in many published results the thought of single phase perovskite materials after low or high temperature treatments is not safe to be assumed, as in ref. [12]. Furthermore, as it is found in our work, the two crystalline phases that may be developed at low sintering temperatures do not deteriorate the electrical properties, but seem to demonstrate broader conductivity versus temperature curves close to maximum value. This may be associated to the theoretical consideration of almost constant charge concentration, as shown in Fig. 9 of Ref. [14]. As can be seen from Fig. 7, as the Mn concentration increases the conductivity increases by shifting the maximum to . This is in agreement to the already published results of ref. [11] who found higher conductivity values as tha Mn content increases too.

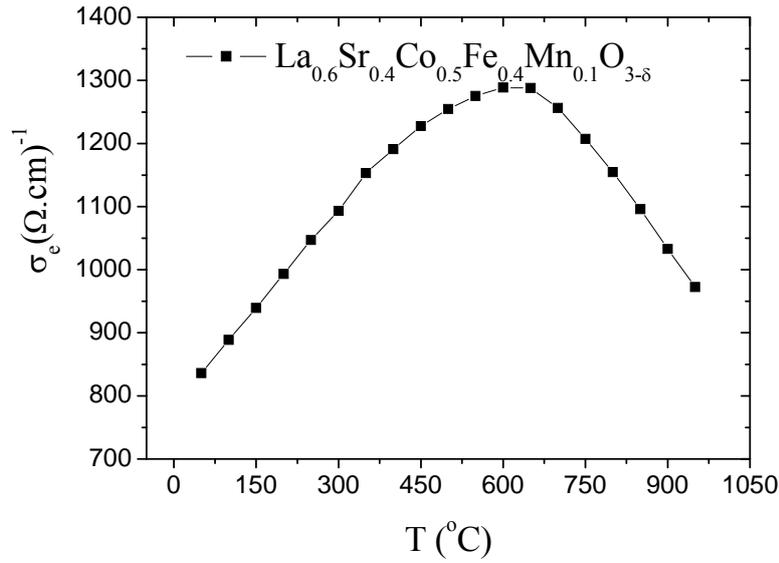

**Figure 7.** Electrical conductivity ($\sigma_e$) of $La_{0.6}Sr_{0.4}Co_{0.5}Fe_{0.4}Mn_{0.1}O_{3-\delta}$ ceramics as a function of temperature.

## 4. Conclusions

In this work, new La-Sr-Co-Fe-Mn-O perovskites have been prepared at 1100 °C by using the well known ceramic method. As it is revealed from the Rietveld analysis, the structural characteristics and phase percentages depend on the La/Sr atomic ratio. The phase mixtures of the composite samples are crystallized in Cubic-Cubic or Cubic-Rhombohedral structures. The microstructure is more or less the same while the grain size is in the range of 0.1 μm to 0.6 μm, i.e. much lower than LSCF at 1100 °C. Microchemistry analysis showed that there is no complete homogenization. As was expected the annealing at higher temperature, as high as 1250 °C, develops larger grains with one perovskite phase. Electrical conductivity measurements suggest that the addition of small amounts of Mn slightly increases the values of $\sigma_e$. We suggest that low temperature sintering reveal two phase system with broader curve maximum and high conductivity for future applications in SOFCs at larger temperature range. Further investigation is needed to examine the effect of doping quantity and the possibly developing a composite system of LSM/LSCF for improving the electrical characteristics as was found in the case of LSCFM perovskites.

**References**

[1] S. Tao, J.T.S. Irvine, J. Electrochem. Soc. 151(2) (2004) A252-259.

[2] K. Tezuka, Y. Hinatsu, A. Nakamura, T. Inami, Y. Shimojo, Y. Morrii, J. Solid State Chem. 141 (1998) 404-410.

[3] S. Xu, Y. Moritomo, K. Ohoyama, A. Nakamura, J. Phys. Soc. Jpn. 72 (2003) 709-712.

[4] M. Verelst, N. Rangavittal, C.N.R. Rao, A. Rousset, J. Solid State Chem. 104 (1993) 74-80.

[5] J. Topfer, J.B. Goodenough, J. Solid State Chem. 130 (1997) 117-128.

[6] L. Morales, R. Zysler, A. Caneiro, Physica B: Cond. Mat. 320 (2002) 100-103.

[7] M. Sahibzada, B.C.H. Steele, K. Zheng, R.A. Rudkin, I.S. Metcalfe, Catalysis Today 38 (1997) 459-466.

[8] H.U. Anderson, L-W. Tai, C.C. Chen, M.M. Nasrallah, W. Huebner, Proceedings of 4th International Symposium On Solid Oxide Fuel Cells (SOFC IV), New Jersey (1995), pp. 375-384.

[9] J.E. ten Elshof, M.H.R. Lankhorst, H.J.M. Bouwmeester, Solid State Ionics 99 (1997) 15-22.

[10] S.P.S. Badwal, T. Bak, S.P. Jiang, J. Love, J. Nowotny, M. Rekas, C.C. Sorrell, E.R. Vance, J. Phys. Chem. Solids 62 (2001) 723-729.

[11] S.M. Plint, P.A. Connor, Sh. Tao, J.T.S. Irvine, Solid State Ionics 177 (2006) 2005-2008.

[12] Q. Xu, D.-P. Huang, W. Chen, J.-H Lee, B.-H. Kim, H. Wang, R.-Z. Yuan, Ceram. Intern. 30 (2004) 429-433.

[13] Scion Image Corporation, http://www.scioncorp.com

[14] J. W. Stevenson, T. R. Armstrong, R. D. Carneim, L. P. Pederson, W. J. Weber, J. Electrochem. Soc. 143 (9) (1996) 2722-2729.